\documentclass[conference]{IEEEtran}
\IEEEoverridecommandlockouts

\usepackage{cite}
\usepackage{amsmath,amssymb,amsfonts}
\usepackage{algorithmic}
\usepackage{graphicx}
\usepackage{textcomp}
\usepackage{xcolor}
\usepackage{url}
\usepackage{multirow}
\usepackage{pifont}%
\newcommand{\cmark}{\ding{51}}%
\newcommand{\xmark}{\ding{55}}%
 \usepackage{booktabs} 
\def\BibTeX{{\rm B\kern-.05em{\sc i\kern-.025em b}\kern-.08em
    T\kern-.1667em\lower.7ex\hbox{E}\kern-.125emX}}
\begin{document}

\title{PURE Codec: Progressive Unfolding of Residual Entropy for Speech Codec Learning
}

\author{Jiatong Shi$^{1}$, Haoran Wang$^{2}$,  William Chen$^{1}$,  Chenda Li$^{2}$,  Wangyou Zhang$^{2}$,  Jinchuan Tian$^{1}$,   Shinji Watanabe$^{1}$ \\
$^{1}$ CMU, $^{2}$ SJTU }



\maketitle

\begin{abstract}
Neural speech codecs have achieved strong performance in low-bitrate compression, but residual vector quantization (RVQ) often suffers from unstable training and ineffective decomposition, limiting reconstruction quality and efficiency. We propose \textbf{PURE Codec} (Progressive Unfolding of Residual Entropy), a novel framework that guides multi-stage quantization using a pre-trained speech enhancement model. The first quantization stage reconstructs low-entropy, denoised speech embeddings, while subsequent stages encode residual high-entropy components. This design improves training stability significantly. Experiments demonstrate that PURE consistently outperforms conventional RVQ-based codecs in reconstruction and downstream speech language model-based text-to-speech, particularly under noisy training conditions.
\end{abstract}


\section{Introduction}

Efficient and high-quality speech coding is critical for a broad range of applications, from real-time communication and mobile telephony to on-device assistants and speech-driven generative models~\cite{cui2024recent, guo2025recent, wu2024towards}. Classical codecs like AMR and Opus~\cite{painter2002perceptual, bessette2003adaptive, opuscodec} rely on hand-engineered features and structured pipelines tailored to speech properties. In contrast, neural speech codecs have emerged as a promising alternative, offering end-to-end learnability, strong rate-distortion performance, and adaptability across conditions~\cite{zeghidour2021soundstream, wu2024codec, shi2024espnet}.

More recently, neural codecs have also played a foundational role in speech language modeling and generative audio synthesis. By producing compact, discrete, and perceptually aligned representations of audio, they support efficient training of large-scale multimodal and multilingual models for speech understanding and generation~\cite{du2024cosyvoice, tian2025espnet, copet2024musicgen}. As such, codec design has become an integral component in the broader landscape of speech and audio foundation models.

Neural codecs typically adopt either \emph{single-stream} or \emph{multi-stream} structures. Single-stream approaches compress all speech information into a unified latent space~\cite{xin2024bigcodec, petermann2021harp, casanova2024lfsc, Guo2024LSCodecLA, messica2024nast, Du2024CosyVoiceAS, langman2024spectral, parker2024scaling, Wu2024TS3CodecTS, ji2024wavtokenizer}, offering simplicity but limited flexibility under varying audio conditions. In contrast, \emph{multi-stream} codecs, often implemented using residual vector quantization (RVQ), employ hierarchical quantization layers to represent different components of the signal~\cite{ai2024apcodec, wu2023audiodec, kumar2023high, Jiang2022DisentangledFL, defossez2022high, gu2024esc, ju2024naturalspeech, du2023funcodec, yang2023hifi, Ahn2024HILCodecHA, yang2024generative, Ji2024LanguageCodecRT, yang2024uniaudio, defossez2024moshi, siuzdak2024snac, zeghidour2021soundstream, zhang2023speechtokenizer}. These methods excel in challenging scenarios such as noisy environments, expressive speech, and low-bitrate settings.

RVQ enables a progressive encoding of residual errors, with each quantization stage refining the approximation left by previous stages~\cite{barnes1996advances}. Recent works have proposed various improvements to RVQ~\cite{gu2024esc, yang2023hifi, siuzdak2024snac}, yet core challenges remain:
\textit{Training instability:} Late-stage quantizers often fail to learn meaningful representations due to poor residual decomposition.
\textit{Redundancy:} Without strong inductive guidance, codebooks may encode overlapping information, undermining compression efficiency.

To overcome these challenges, we propose \textbf{PURE Codec}, a new codec architecture that introduces \emph{enhancement-aware supervision} into the RVQ pipeline. By leveraging a pre-trained denoising model, the first quantization stage is trained to reconstruct clean, low-entropy representations of speech. Subsequent layers are then progressively tasked with encoding residual entropy.\footnote{While enhancement supervision is applied only to the first stage, the residual vector quantization process itself unfolds progressively across the RVQ. The enhanced signal serves to anchor this unfolding in a structured and perceptually informed manner.} This design improves decomposition quality and stabilizes training.\footnote{The codebase is shared at \scriptsize{https://github.com/ftshijt/espnet/tree/se\_dac} and model checkpoints are released at \scriptsize{https://huggingface.co/collections/espnet/neural-codecs-67cb8c96859c53a6131a85ec}.}

The key contributions of this work are: (1) A novel RVQ-based codec guided by a speech enhancement model to anchor low-entropy reconstruction in the early quantization stage. (2) A progressive entropy decomposition scheme that yields efficient representations across varying bitrate regimes. The training is stabilized by combining a variational auto-encoder (VAE) initialization and a stochastic enhancement scheduler to balance structure and diversity during optimization. (3) Empirical evidence demonstrating superior performance in terms of reconstruction fidelity, robustness, and flexibility compared to existing neural codecs.

\section{Background}
\label{sec:background}

\subsection{Quantization and Residual Vector Quantization}

Neural speech codecs aim to convert continuous acoustic signals into compact, discrete representations for efficient transmission and storage. A typical codec architecture comprises an encoder $\mathrm{Enc}(\cdot)$, a quantizer $\mathrm{Quant}(\cdot)$, and a decoder $\mathrm{Dec}(\cdot)$. Given an input waveform $S \in \mathbb{R}^{1 \times T_S}$ with the speech length as $T_S$, the encoder produces a sequence of frame-level embeddings $Q = \mathrm{Enc}(S) \in \mathbb{R}^{D \times T}$, where $D$ is the feature dimension and $T$ is the number of frames.

To discretize these embeddings, many neural codecs employ RVQ as the quantizer, which encodes the signal progressively through a sequence of quantization stages. Specifically, RVQ uses $L$ codebooks $\{\mathcal{B}^1, \dots, \mathcal{B}^L\}$, each containing $B$ vectors in $\mathbb{R}^D$. Here, the frame-level embedding sequence $Q$ can be elaborated into a sequence of embedding states $[q_1, \dots, q_T]$. At each frame $t$, quantization begins with the raw embedding $r_t^0=q_t$ and iteratively selects the nearest codeword embedding from each codebook to reduce the residual error:
\begin{align}
\label{eq:rvq}
c_t^l &= \arg\min_j \| r_t^{l-1} - b_j^l \|_2^2 \in \{1, ..., B\}, \\
\hat{q}_t^{(l)} & = b_{c_t^l}^l, \\
r_t^l &= r_t^{l-1} - b_{c_t^l}^l,
\label{eq:rvq2}
\end{align}
where $c_t^l$ is the selected index at the stream $l$ and $b_{c_t^l}^l$ is the corresponding codeword embedding. The quantized embedding $\hat{q}_t$ for frame $t$ is the sum of selected vectors across all streams:
\begin{equation}
    \hat{q}_t = \sum_{l=1}^L \hat{q}_t^{(l)} = \sum_{l=1}^L b_{c_t^l}^l.
\end{equation}
The decoder reconstructs the waveform from the sequence $\hat{Q} = [\hat{q}_1, \dots, \hat{q}_T]$ via $\hat{S} = \mathrm{Dec}(\hat{Q})$. The whole quantization process can be defined as 
\begin{equation}
    \left(\hat{Q}, \{c_t^l\}_{t=1,\,l=1}^{T,\,L}\right) = \mathrm{Quant}(Q),
\end{equation}

\subsection{Extensions and Challenges of RVQ in Neural Codecs}

While RVQ offers a natural multi-stage structure for neural quantization, its performance in practice is often limited by training instability and poor decomposition of information across layers. Recent studies have proposed various extensions to enhance RVQ, such as cross-scale RVQ~\cite{gu2024esc}, group RVQ~\cite{yang2023hifi}, and multi-scale RVQ~\cite{siuzdak2024snac}. These methods aim to improve both the representational capacity and optimization behavior of RVQ-based codecs. 

While these approaches have led to improvements in expressivity and bitrate flexibility, they often rely on architectural modifications or complex training heuristics without explicitly aligning quantization stages with the underlying entropy structure of the speech signal. As a result, stage-level redundancy and training instability remain open challenges, especially in low-bitrate or noisy training conditions. These challenges motivate a reformulation of RVQ that directly guides the decomposition process using signal structure cues, a gap we address in our proposed PURE Codec framework.

\subsection{Motivation for Enhancement-Guided Quantization}

To address these limitations, we propose a reformulation of residual vector quantization (RVQ) that explicitly aligns the quantization stages with the natural information structure of speech. Our key insight is that enhanced (denoised) speech signals, produced by high-quality speech enhancement models, exhibit lower entropy than their original noisy counterparts.\footnote{In our hypothesis, the clean signal is more informative compared to the noisy signal and can be regarded as lower entropy, which is further confirmed in the following perceptual entropy analysis.} These enhanced signals effectively suppress stochastic noise and background interference while preserving the underlying phonetic and prosodic structure, resulting in more regular, less variable, and inherently more compressible representations.

By guiding the first quantization stage to reconstruct these low-entropy embeddings, we encourage the early codebooks to focus on structured, perceptually salient components of the signal. This design allows subsequent stages to naturally refine the residual, higher-entropy content. The resulting quantization hierarchy progressively unfolds the residual entropy of speech, stabilizing training, reducing redundancy, and yielding a more modular codec architecture.

This design principle is further supported by our analysis of perceptual entropy (PE), a psychoacoustic measure estimating the theoretical lower bound of perceptual coding rate~\cite{johnston1988estimation, painter2002perceptual, shi2021sequence}. Using 5,000 randomly sampled utterances from the URGENT Challenge training set~\cite{zhang2024urgent}, we compute PE for both noisy and enhanced speech.\footnote{We use the implementation of \cite{shi2021sequence} for PE calculation.} Results show that enhancement reduces PE by 57.80\% on average, confirming that enhanced speech is substantially more compressible from a perceptual standpoint. This observation underscores the utility of using enhancement-guided targets in the quantization process.

This principle forms the foundation of our proposed framework, PURE Codec, which integrates enhancement-aware supervision into RVQ and decomposes speech entropy in a structured, progressive manner. In the following section, we detail the architecture and training methodology of PURE Codec, and demonstrate how this design enables efficient, scalable, and robust speech compression.

\begin{figure}
    \centering
    \includegraphics[width=\linewidth]{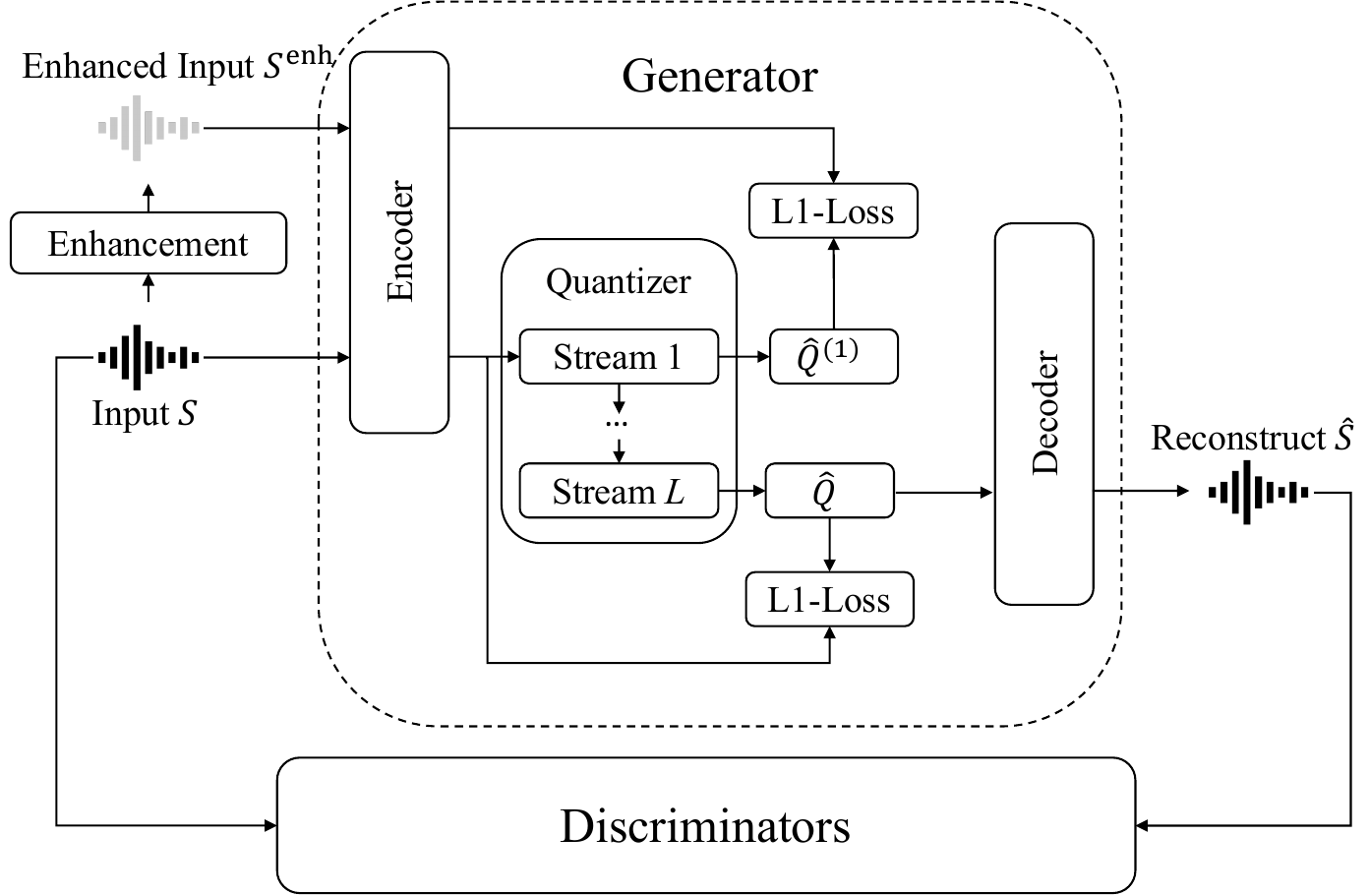}
    \caption{PURE Codec framework. The input waveform $S$ is optionally enhanced to produce $S^{\text{enh}}$, guiding the first quantization stage via low-entropy embeddings. The encoder output is refined through multiple quantization streams and decoded to reconstruct $\hat{S}$. L1 losses and adversarial losses from discriminators supervise training. Refer to Sec.~\ref{sec: purecodec} for details.}
    \label{fig:overall}
    \vspace{-15pt}
\end{figure}

\section{PURE Codec}
\label{sec: purecodec}

\subsection{General Framework}

PURE Codec is a multi-stage, vector-quantized neural speech codec that introduces a residual decomposition aligned with the entropy profile of the input signal. Building on the RVQ framework, PURE incorporates a novel enhancement-guided hierarchy, where early stages (Stream 1) encode enhanced (low-entropy) content, and later stages (up to Stream $L$) progressively capture noisier (higher-entropy) residuals.

As illustrated in Figure~\ref{fig:overall},  the codec follows the basic design of the RVQ-based codec discussed in Sec.~\ref{sec:background}.
In parallel with the basic encoder-decoder design, an optional enhancement module processes $S$ to produce a denoised waveform $S^{\text{enh}} = \mathrm{Enh}(S)$, which is then encoded to produce an enhanced embedding $\tilde{Q} = \mathrm{Enc}(S^{\text{enh}})$. This enhanced embedding serves as a low-entropy anchor for supervising the first quantization stream. Each stream in the quantizer sequentially refines the approximation of $Q$ using residual codebook lookups. The first stream is explicitly trained to approximate $\tilde{Q}$, while the following streams (2 to $L$) model the residuals.\footnote{Please refer Sec.~\ref{ssec: pure} for more details.} The quantized embedding $\hat{Q} = [\hat{q}_1, \dots, \hat{q}_T]$ is then passed through the decoder to reconstruct the waveform $\hat{S} = \mathrm{Dec}(E)$.

\subsection{Progressive Unfolding of Residual Entropy}
\label{ssec: pure}



We formalize the PURE Codec’s entropy-guided quantization as follows. Let $\tilde{q}_t = \mathrm{Enc}(S^{\text{enh}})_t$ be the enhanced embedding at frame $t$, and $q_t = \mathrm{Enc}(S)_t$ the original embedding (following the same formulation as Sec.~\ref{sec:background}). The first quantization stage minimizes the error between $\tilde{q}_t$ and its closest codebook entry:
\begin{align}
c_t^1 &= \arg\min_j | \tilde{q}_t - b_j^1 |_2^2 \in \{1, \dots B\}, \\
\hat{q}_t^{(1)} &= b_{c_t^1}^1, \\
r_t^1 & = q_t - \hat{q}_t^{(1)}.
\end{align}
We denote $\hat{Q}^{(1)} = [\hat{q}_1^{(1)}, \dots, \hat{q}_T^{(1)}]$, corresponding to the quantized embedding shown in Figure~\ref{fig:overall}. The residual $r_t^1$ is then passed to higher stages, each iteratively refining the approximation following the same mechanism as Eq.~\eqref{eq:rvq}-\eqref{eq:rvq2}. The final quantized embedding is the sum over all stages:
\begin{equation}
\hat{q}_t = \sum_{l=1}^L \hat{q}_t^{(l)}, 
\end{equation}
where the sequence $\hat{Q}$ is defined as $[\hat{q}_1, ..., \hat{q}_T]$.

This progressive unfolding of residual entropy aligns each quantization stage with a specific portion of the signal's information structure. 
As a result, the system produces more controllable representations, and it can gracefully adapt to varying bitrate constraints through early stopping of the quantization hierarchy.

\begin{table*}[t]
\centering
\caption{Overview of PURE Codec variant configurations described in Section~\ref{ssec:setup}. 
Each model explores different enhancement setups.
\textbf{Enhancement Model} refers to the pre-trained model from Hugging Face.
\textbf{Delayed Intro} indicates if the enhancement is introduced later in training.
\textbf{Multi-stream} enables an additional stream for enhanced speech reconstruction.
\(\boldsymbol{p_{\text{enh}}}\) is the probability of using enhancement defined in Eq.~\eqref{eq:penh}.
\textbf{Encoder Fine-tune} specifies whether the encoder is fine-tuned.
}
\vspace{-10pt}
\label{tab:model_variants}
\resizebox{0.9\linewidth}{!}{
\begin{tabular}{l|lcc|cccc}
\toprule
\textbf{Model Tag} & \textbf{Enhancement Model (HF Tag)} & \textbf{\#Param} & \textbf{Training data} & \textbf{Delayed Intro} & \textbf{Multi-stream} & \(\boldsymbol{p_{\text{enh}}}\) & \textbf{Encoder Fine-tune} \\
\midrule
PURE & wyz/tfgridnet\_for\_urgent24 & 8.5 M & $>$1000 h & \multirow{4}{*}{\xmark}  & \multirow{4}{*}{\xmark}  & \multirow{4}{*}{0.50} & \multirow{4}{*}{\xmark} \\
Abl.A & wyz/vctk\_dns2020\_whamr\_bsrnn\_large\_noncausal & 63.1 M & 157 h \\
Abl.B & wyz/vctk\_dns2020\_whamr\_bsrnn\_medium\_noncausal & 16.9 M & 157 h &  &  &  \\
Abl.C & wyz/vctk\_dns2020\_whamr\_bsrnn\_small\_noncausal & 4.8 M & 157 h & & \\
\midrule
Abl.D & \multirow{5}{*}{wyz/tfgridnet\_for\_urgent24} & \multirow{5}{*}{8.5 M} & \multirow{5}{*}{$>$1000 h} & \cmark & \xmark  & 0.50 & \xmark \\
Abl.E &  &  &  & \xmark  & \cmark & 0.50 & \xmark \\
Abl.F &  &  &  & \xmark  & \xmark  & 0.25 & \xmark \\
Abl.G &  &  &  & \xmark  & \xmark  & 0.75 & \xmark \\
Abl.H &  &  &  & \xmark  & \xmark  & 0.50 & \cmark \\

\bottomrule
\end{tabular}
}
\vspace{-15pt}
\end{table*}

\subsection{Training Strategy}

PURE Codec is trained in two stages. First, we pretrain the encoder-decoder pair as a VAE to ensure that they learn a smooth and expressive latent space. Then, we introduce quantization and enhancement-guided supervision.

In the VAE pretraining stage, the encoder and decoder are trained to minimize a combination of waveform reconstruction loss and a KL divergence regularization term. Specifically, the loss includes an $\ell_1$ waveform loss, a multi-resolution mel loss, and a KL divergence penalty (i.e., $D_{\mathrm{KL}}$):
\begin{equation}
\mathcal{L}_{\text{VAE}} = \| S - \hat{S} \|_1 + \mathrm{MelLoss}(S, \hat{S}) + \lambda_{\text{KL}} \, D_{\mathrm{KL}}(q(z|S) \| p(z)),
\end{equation}
where $z$ is the latent variable drawn from the approximate posterior, and $p(z)$ is the standard Gaussian prior, $\lambda_{\text{KL}}$ is a hyperparameter to control the KL loss. This step allows the model to learn the underlying structure of speech without discrete bottlenecks, which facilitates stable convergence in the subsequent stage.

After VAE pretraining, we introduce the quantization layers and incorporate enhancement supervision. The key design here is a stochastic scheduling mechanism that determines whether the first quantization stage should align with the enhanced embedding or the original embedding. Specifically, with a fixed probability $p_{\text{enh}}$, we use the enhanced embedding $\tilde{Q} = \mathrm{Enc}(\mathrm{Enh}(S))$ as the target for the first-stage quantization; otherwise, we use $Q = \mathrm{Enc}(S)$. This stochastic scheduling helps balance robustness and flexibility during training. The corresponding loss is:
\begin{equation}
\label{eq:penh}
\mathcal{L}_{\text{enh}} = \mathbb{E}_{\mathrm{Bernoulli}(p_{\text{enh}})} \left[ \| \hat{Q}^{(1)} - \tilde{Q} \|_2^2 \right].
\end{equation}

The full training of PURE Codec is framed within a GAN-based modeling paradigm, where the codec acts as the generator \( \mathcal{G} \), comprising the encoder, quantizer, and decoder modules. A multi-scale discriminator \( \mathcal{D} \) is trained to distinguish real waveforms from those synthesized by \( \mathcal{G} \).

The reconstruction loss promotes fidelity at both waveform and perceptual levels:
\begin{equation}
\mathcal{L}_{\text{rec}} = \| S - \hat{S} \|_1 + \mathrm{MelLoss}(S, \hat{S}) +  \mathrm{MelLoss}(\mathrm{Dec}(\mathrm{Enc}(S)), S),
\end{equation}
where $\mathrm{MelLoss}$ denotes a multi-resolution mel-spectrogram loss. The vector quantization loss~\cite{van2017neural} is defined as:
\begin{equation}
\mathcal{L}_{\text{vq}} = \sum_{l=1}^L \left( \| \mathrm{sg}[Q] - M^l \|_2^2 + \beta \| Q - \mathrm{sg}[M^l] \|_2^2 \right),
\end{equation}
with stop-gradient operator $\mathrm{sg}[\cdot]$, codebook output $M^l=[\sum_{k=1}^l\hat{q}_1^(k), \dots, \sum_{k=1}^l\hat{q}_T^(k)]$ at stage $l$, and a hyperparameter $\beta$. The adversarial generator loss follows the basic GAN formulation:
\begin{equation}
\mathcal{L}_{\text{adv}}^{\mathcal{G}} = \mathbb{E}_{\hat{S}} \left[ (\mathcal{D}(\hat{S}) - 1)^2 \right] ,
\end{equation}
which encourages the synthesized waveform $\hat{S}$ to be indistinguishable from the real waveform under the discriminator’s judgment. The generator \( \mathcal{G} \) is optimized using a composite loss:
\begin{equation}
\label{eq:g}
\mathcal{L}_{\mathcal{G}} = \lambda_{\text{enh}} \mathcal{L}_{\text{enh}} + \lambda_{\text{rec}} \mathcal{L}_{\text{rec}} + \lambda_{\text{vq}} \mathcal{L}_{\text{vq}} + \lambda_{\text{adv}} \mathcal{L}_{\text{adv}}^{\mathcal{G}},
\end{equation}
where $\lambda_{\text{enh}}$, $\lambda_{\text{rec}}$, $\lambda_{\text{vq}}$, and $\lambda_{\text{adv}}$ are hyperparameters for losses.

Concurrently, the discriminator \( \mathcal{D} \) is trained to distinguish real and generated samples:
\begin{equation}
\label{eq:d}
\mathcal{L}_{\mathcal{D}} = \mathbb{E}_{S} \left[ (\mathcal{D}(S) - 1)^2 \right] + \mathbb{E}_{\hat{S}} \left[ (\mathcal{D}(\hat{S}))^2 \right] + \mathcal{L}_{\text{feat}}(S, \hat{S}),
\end{equation}
where $\mathcal{L}_{\text{feat}}(S, \hat{S})$ is the feature matching loss.

The full training process of PURE Codec can be expressed as the following min-max optimization:
\begin{equation}
\min_{\mathcal{G}} \max_{\mathcal{D}} \; \mathcal{L}_{\text{PURE}}(\mathcal{G}, \mathcal{D}) = \mathcal{L}_{\mathcal{G}} - \lambda_{\text{adv}} \cdot \mathcal{L}_{\mathcal{D}}.
\end{equation}

\section{Experiments}

\subsection{Dataset}

We evaluate PURE Codec across three representative datasets to cover a wide range of speech conditions. The first dataset, \textbf{OWSM-v3.2}~\cite{tian2024effects}, is used for training and represents a general-purpose, multi-domain speech corpus that includes diverse speaking styles, accents, and acoustic environments. The second dataset, \textbf{CommonVoice} (V13)~\cite{ardila2020common},\footnote{We combine training subsets in CommonVoice that have at least an hour.} consists of crowdsourced recordings that exhibit a wider range of speaker variability and lower recording quality, serving as a benchmark for low-quality input scenarios. Finally, we include \textbf{URGENT} 2024 training set~\cite{zhang2024urgent},\footnote{The URGENT dataset is a simulated noisy corpus derived from five public English sources: DNS5 LibriVox~\cite{dubey2024icassp}, LibriTTS~\cite{zen2019libritts}, CommonVoice 11.0~\cite{ardila2020common}, VCTK, and WSJ~\cite{paul1992design}. Noise sources are drawn from DNS5 challenge data, originally collected from Audioset~\cite{gemmeke2017audio} and Freesound~\cite{fonseca2017freesound}, and from WHAM!~\cite{wichern2019wham}. Reverberation is simulated using room impulse responses generated by the image method~\cite{dubey2024icassp}. We follow the same configuration as in~\cite{zhang2024urgent}: signal-to-noise ratios are sampled from $\mathcal{U}(-5, 20)$\,dB; reverberation and clipping are applied with 25\% probability; and bandwidth limitations are randomly selected from \{8, 16, 22.05, 24, 32\}\,kHz.} a simulated dataset with artificially injected environmental noise and reverberation, to test robustness under extremely noisy training conditions. All three training sets are finally downsampled to a 16 kHz sampling rate. The final data sizes are 150k hours, 16k hours, and 2k hours for the three training sets.\footnote{Due to chunk-based training with fixed batch size and step count, the effective training exposure is approximately 40k hours. To match this constraint, we randomly subsample the OWSM-v3.2 corpus to 40k hours.}

\begin{table*}[t]
\centering
\caption{
Training stability analysis with different training datasets.
Evaluation results on resynthesized \textbf{Librispeech-test-clean} speech across reconstruction metrics. 
Note that the original DAC, corresponding to the open-source DAC model in its original work, is shown for reference only due to substantial differences in training setup and data.
}
\vspace{-5pt}
\label{tab:codec-results-overall}
\resizebox{0.9\textwidth}{!}{
\begin{tabular}{ll|
c c c c c c c}
\toprule
\textbf{Training Data} & \textbf{Model} 
& \textbf{SDR $\uparrow$} 
& \textbf{PESQ $\uparrow$} 
& \textbf{UTMOS $\uparrow$} 
& \textbf{DNSMOS $\uparrow$} 
& \textbf{VISQOL $\uparrow$} 
& \textbf{WER $\downarrow$} 
& \textbf{SPK-SIM $\uparrow$} \\
\midrule
-- & Ground Truth & -- & -- & 4.09 & 3.18 & -- & 2.73 & -- \\
-- & Original DAC & 1.79 & 3.40 & 3.60 & 3.16 & 4.49 & 2.68 & 0.73 \\
\midrule
\multirow{2}{*}{OWSM-v3.2 (150k hours)} & Baseline DAC & \textbf{4.01} & 2.37 & 3.42 & 3.17 & 4.34 & 2.26 & 0.64 \\
& PURE Codec   & 2.17 & 2.62 & \textbf{3.64} & \textbf{3.21} & \textbf{4.41} & \textbf{2.05} & 0.71 \\
\midrule
\multirow{2}{*}{Commonvoice (16k hours)} & Baseline DAC & -5.21  & 1.36 & 1.45 & 2.13 & 4.07 & 3.00 & 0.31 \\
& PURE Codec   & 2.70  & \textbf{2.70} & 3.65 & 3.18 & 4.39 & 2.14 & \textbf{0.76} \\
\midrule
\multirow{2}{*}{URGENT (2k hours)} & Baseline DAC & -6.79  & 1.32 & 1.31 & 1.97 & 4.12 & 3.67 & 0.38 \\
 & PURE Codec   & 1.35  & 2.50 & 3.41 & 3.09 & 4.36 & 2.07 & 0.54 \\
\bottomrule
\end{tabular}
}
\vspace{-10pt}
\end{table*}

\begin{table*}[t]
\centering
\caption{
\textbf{Training Set: OWSM-V3.2.} 
Evaluation results on resynthesized \textbf{Librispeech-test-clean} speech across reconstruction metrics. 
}
\vspace{-5pt}
\label{tab:codec-results-ablation}
\resizebox{0.8\textwidth}{!}{
\begin{tabular}{l|
c c c c c c c c}
\toprule
\textbf{Model} 
& \textbf{SDR $\uparrow$} 
& \textbf{PESQ $\uparrow$} 
& \textbf{UTMOS $\uparrow$} 
& \textbf{DNSMOS $\uparrow$} 
& \textbf{VISQOL $\uparrow$} 
& \textbf{WER $\downarrow$} 
& \textbf{SPK-SIM $\uparrow$} \\
\midrule
Ground Truth & -- & -- & 4.09 & 3.18 & -- & 2.73 & -- \\

\midrule
Baseline DAC & 4.01 & 2.37 & 3.42 & 3.17 & 4.34 & 2.26 & 0.64 \\
\midrule
PURE Codec   & 2.17 & 2.62 & 3.64 & 3.21 & 4.41 & \textbf{2.05} & 0.71 \\
Abl.A        & 2.35 & 2.64 & 3.66 & 3.19 & \textbf{4.42} & 2.10 & 0.73 \\
Abl.B        & 2.81 & 2.63 & 3.66 & 3.20 & 4.35 & 2.42 & 0.73 \\
Abl.C        & 2.86 & 2.66 & 3.62 & 3.20 & 4.36 & 2.21 & 0.72 \\
Abl.D        & 2.59 & 2.69 & 3.66 & 3.19 & 4.32 & 2.09 & 0.72 \\
Abl.E        & 2.80 & 2.35 & 3.32 & 3.12 & 4.26 & 2.24 & 0.70 \\
Abl.F        & \textbf{3.00} & \textbf{2.85} & \textbf{3.79} & \textbf{3.24} & 4.26 & 2.21 & 0.77 \\
Abl.G        & 2.48 & 2.66 & 3.65 & 3.18 & 4.41 & 2.13 & \textbf{0.78} \\
Abl.H        & 0.74 & 2.23 & 3.23 & 3.16 & 4.25 & 2.56 & 0.59 \\
\bottomrule
\end{tabular}
}
\vspace{-5pt}
\end{table*}

\subsection{Experimental Setup}
\label{ssec:setup}

We use ESPnet-Codec as the training framework~\cite{shi2024espnet}.

\noindent \textbf{Baseline Setup}. As a baseline for comparison, we adopt a DAC-style codec~\cite{kumar2023high}. The generator \(\mathcal{G}\) consists of an encoder, a multi-stage quantizer, and a decoder, with a $D=512$-dimensional hidden state and codebook size. We use $L=8$ vector quantizers with $B=1024$ bins, initialized using K-means. Quantization dropout is used to simulate various target bitrates (0.5, 1, 2, and 4 kbps). The discriminator \(\mathcal{D}\), defined in Eq.~\eqref{eq:d}, follows a multi-scale, multi-period, and multi-band design, combining fast Fourier transform (FFT) based and periodic modules to provide fine-grained adversarial supervision across both time and frequency domains.

The model is optimized with a combination of waveform reconstruction, adversarial, and vector quantization losses. Each loss is weighted by hyperparameters: \(\lambda_{\text{rec}} = 1.0\), \(\lambda_{\text{vq}} = 0.25\), and \(\lambda_{\text{adv}} = 1.0\) as defined in Eq~\eqref{eq:g}. Optimization uses AdamW with a learning rate of \(2 \times 10^{-4}\) and exponential decay (\(\gamma = 0.999875\)). Training runs for 360 epochs (3k steps per epoch) using a batch size of 64 and chunked waveform segments of 32{,}000 samples. The best-performing checkpoints are selected based on validation metrics.\footnote{For more detailed hyperparameters that are not specified, we follow the same DAC setting as the ESPnet AMUSE recipe~\cite{shi2024espnet}.}

In addition to our retrained DAC baselines on each training set, we also report results from the original DAC model using publicly available weights. These results are provided for reference only, as the original DAC was trained on a different dataset and operates at a higher bitrate (6 kbps) compared to the 4 kbps setting used in our experiments. This mismatch in training data and bitrate should be considered when interpreting performance comparisons.

\noindent \textbf{PURE Codec Setups.} To ensure a fair comparison, PURE Codec adopts the same backbone architecture as the DAC baseline. The training is performed in two stages. In the first stage, we pre-train the encoder-decoder pair as a variational autoencoder (VAE) for 180 epochs (3{,}000 steps per epoch) using a batch size of 64. In the second stage, we incorporate enhancement-guided supervision using a pre-trained TFGridNet model, originally developed for the URGENT 2024 Challenge~\cite{zhang2024urgent, wang2023tf}.\footnote{https://huggingface.co/wyz/tfgridnet\_for\_urgent24} The enhancement sampling probability \(p_{\text{enh}}\) is set to 0.5 in Eq.~\eqref{eq:penh}. During this stage, we use a batch size of 32 and freeze the encoder to stabilize training.

To further investigate the effects of different design choices, we explore several PURE Codec variants by modifying the enhancement model~\cite{zhang2024scale}, introducing the enhancement module at different training stages, enabling joint fine-tuning of the encoder, employing multi-stream modeling for enhanced speech, and varying the value of \(p_{\text{enh}}\). Detailed configurations of each variant are summarized in Table~\ref{tab:model_variants}.

To further investigate the impact of key design decisions in PURE Codec, we conduct ablation studies across several controlled variants. These variants differ in terms of enhancement model architecture, training schedule, encoder update strategy, and enhancement supervision probability. The detailed configurations for each variant are summarized in Table~\ref{tab:model_variants}. (1) Enhancement Model Scaling (Abl.A--Abl.C). We vary the enhancement frontend by replacing the default TF-GridNet~\cite{zhang2024scale} with an alternative model, BSRNN, and its smaller and larger variants. This allows us to assess how model capacity and architecture influence quantization behavior and reconstruction quality.
(2) Enhancement Injection Timing (Abl.D). We investigate when to introduce enhancement supervision during training. Starting with a VAE, we first add quantization, then enable enhancement at different points after 50k steps to evaluate the effect of supervision timing on training dynamics and entropy decomposition. 
(3) Multi-stream enhancement supervision (Abl. E). In the default setting, the first quantization stream is guided by enhancement supervision. In this variant, we instead apply supervision to the second stream, aiming to evaluate whether shifting enhancement guidance to a later stream can better capture high-fidelity enhanced features.
 (3) Encoder Fine-tuning (Abl.H).
By default, the encoder is frozen after VAE pretraining. We test a variant that allows joint fine-tuning of the encoder, quantizer, and decoder to study the trade-off between stability and adaptability.
(4) Enhancement Supervision Probability (Abl.F--Abl.G).
We vary the probability \(p_{\text{enh}}\) of applying enhancement supervision during training. Beyond the default setting of 0.5, we evaluate 0.25 and 0.75 to examine how frequent guidance affects entropy regularization and overall performance.

\subsection{Evaluation}\label{sec:eval}

The evaluation is conducted using the VERSA toolkit~\cite{shi2024versa}. We report the following metrics: signal-to-distortion ratio~(SDR) is used to assess waveform-level reconstruction quality. For perceptual quality evaluation, we include the Perceptual Evaluation of Speech Quality (PESQ)~\cite{painter2000perceptual}; UTMOS from VoiceMOS2022 challenge~\cite{1360861705599880960}, a neural mean opinion score (MOS) estimator trained on human ratings; and the deep noise suppression MOS following ITU-T P.835~(DNSMOS)~\cite{reddy2022dnsmos}, which estimates overall quality in noisy conditions. We also report the virtual speech quality objective listener (VISQOL)~\cite{hines2015visqol}, which captures spectral and perceptual similarity based on frequency-domain analysis. For application-level evaluation, we compute word error rate~(WER) using Whisper-large-V3~\cite{radford2023robust} to assess linguistic intelligibility, and speaker similarity (SPK-SIM), which measures cosine similarity between speaker embeddings extracted from reference and resynthesized audio to quantify speaker identity preservation.\footnote{We use the RawNet3-based speaker embedding extractor based on ESPnet-SPK~\cite{jung2024espnet}.} All metrics are evaluated on the Librispeech-test-clean set~\cite{7178964}. Higher values indicate better performance for all metrics except WER.

\subsection{Main Results and Training Stability}
\label{ssec:main-results}

We first compare the PURE Codec with the DAC-style baseline under multiple training datasets (OWSM-v3.2, CommonVoice, and URGENT), as shown in Tables~\ref{tab:codec-results-overall}. Across all conditions, PURE Codec demonstrates consistently improved reconstruction and application-level performance, implying that our method is general and non-sensitive across datasets.

On the OWSM-v3.2 setting, PURE improves WER from 2.26\% to 2.05\%, UTMOS from 3.42 to 3.64, and speaker similarity from 0.64 to 0.71. These results indicate that PURE enhances both intelligibility and perceptual quality, even when the SDR is lower, likely due to codec-induced amplitude shifts that do not significantly affect perceptual metrics.

More importantly, we observe that training instability, a common issue in RVQ-based codecs, is substantially reduced in PURE. When training the DAC baseline on more challenging corpora such as CommonVoice and URGENT, we find significant degradation or even collapse, with SDR values dropping to as low as -6.79 and PESQ below 1.4. In contrast, PURE remains stable across all datasets, achieving SDR over 1.3 and PESQ over 2.5 even under highly noisy training conditions. This confirms the hypothesis that entropy-guided quantization anchored by enhancement supervision stabilizes training and improves robustness.

\subsection{Ablation Study}
\label{ssec:ablation-results}

We conduct a comprehensive ablation study (Table~\ref{tab:codec-results-ablation}) to examine the design choices underlying PURE Codec. The results highlight several key insights:

(1) Enhancement Model Scaling (Abl.A–C). Replacing the default TF-GridNet with smaller or alternative enhancement frontends (e.g., BSRNN) shows modest differences. Larger models (Abl.C) slightly improve PESQ and SDR, suggesting that while enhancement quality helps, the core benefits of PURE come from the structural alignment rather than the specific enhancement model.
(2) Enhancement Injection Timing (Abl.D). Delayed introduction of enhancement guidance stabilizes early optimization but results in a slight trade-off between perceptual quality and intelligibility. This indicates the importance of timely supervision to shape early-stage quantizer behavior.
(3) Enhancement Supervision Frequency (Abl.F/G).
The sampling probability $p_{\text{enh}}$ in Eq.~\eqref{eq:penh} plays a critical role. Abl.F, with $p_{\text{enh}}=0.25$, achieves the best performance overall (e.g., PESQ = 2.85, UTMOS = 3.79), suggesting that sparse but consistent guidance helps balance compression flexibility and entropy control.
(4) Multi-stream Extension (Abl.E). Adding a separate stream for enhanced speech reconstruction does not improve performance and slightly hurts perceptual metrics, likely due to optimization interference between streams.
(5) Encoder Fine-tuning (Abl.H).
Allowing encoder updates after VAE pretraining leads to catastrophic degradation (e.g., SDR drops to 0.74), reaffirming that encoder freezing is essential for training stability in PURE.

\begin{table}[t]
\centering
\caption{SpeechLM-based TTS performance on LibriSpeech. UTMOS reflects perceptual quality, WER measures intelligibility, and SPK-SIM quantifies speaker similarity.}
\label{tab:speechlm-tts}
\vspace{-5pt}
\resizebox{0.9\linewidth}{!}{
\begin{tabular}{l|ccc}
\toprule
\textbf{Codec}& \textbf{WER $\downarrow$}  & \textbf{SPK-SIM $\uparrow$} & \textbf{UTMOS $\uparrow$}  \\
\midrule
Baseline DAC & 10.8 & 0.70 & 3.68 \\
\midrule
PURE Codec & \textbf{10.5} & 0.68 & 3.95 \\
Abl.A & 14.0 & 0.70 & 3.92 \\
Abl.B & 11.1 & \textbf{0.71} & \textbf{3.99} \\
Abl.C & 11.6 & 0.69 & 3.96 \\
\bottomrule
\end{tabular}
}
\vspace{-5pt}
\end{table}

\subsection{Performance in SpeechLM-based TTS}

To assess downstream usability in generation tasks, we integrate the proposed codec into a SpeechLM-based text-to-speech (TTS) system. Specifically, we train a decoder-only conditional SpeechLM on LibriSpeech-960h, where the codec tokens serve as the input speaker prompt and the target for synthesis. We use the delay-interleave pattern to shift each token stream \cite{copet2024musicgen} for better acoustic modelling. All models are trained for 400K steps, while inference is performed with top-k sampling, with $k=30$ and a sampling temperature of 0.7. Generated samples are evaluated using 3 proxy metrics: WER for intelligibility, UTMOS for audio, and embedding distance for SPK-SIM, with the implementations described in Section \ref{sec:eval}.
Table~\ref{tab:speechlm-tts} summarizes the results for Baseline DAC, PURE Codec, and Ablation variants A–C.

We observe that the PURE Codec consistently improves downstream generation across all metrics. Compared to DAC, PURE reduces WER from 2.91\% to 2.46\%, indicating better linguistic consistency. UTMOS increases to 3.78, and SPK-SIM reaches 0.76, reflecting improved perceptual and speaker quality. Ablation results suggest that the benefits of enhancement-guided quantization hold across different enhancement backbones, with Abl.C performing comparably to the full PURE model. These findings highlight the importance of entropy-aware quantization for speech LM-based generation, which lead to more reliable synthesis.

\section{Conclusion}

We presented \textbf{PURE Codec}, a speech coding framework that improves RVQ by progressively unfolding residual entropy through enhancement-guided supervision. By anchoring the first quantization stage to low-entropy, denoised embeddings, PURE achieves more stable training quantization. Our two-stage training strategy, combining variational pretraining and stochastic enhancement scheduling, leads to consistent improvements over baselines across clean, noisy, and low-quality speech conditions.

A current limitation is that PURE relies on speech-specific enhancement models and is not directly applicable to general audio. Future work will explore extending this entropy-guided quantization paradigm to broader audio domains.

\section*{Acknowledgment}

This work used the Bridges2 at PSC and Delta/DeltaAI NCSA systems through CIS210014 from the ACCESS program, supported by NSF \#2138259, \#2138286, \#2138307, \#2137603, and \#2138296.


\bibliographystyle{IEEEbib}
\bibliography{ref}

@string{icassp = "Proc. ICASSP"}

@string{interspeech = "Proc. Interspeech"}

@string{neurips = "Proc. NeurIPS"}

@string{waspaa = "Proc. WASPAA"}

@string{emnlp = "Proc. EMNLP"}

@string{slt = "Proc. SLT"}

@string{ismir = "Proc. ISMIR"}

@string{iclr = "Proc. ICLR"}

@string{icml = "Proc. ICML"}

@string{acl = "Proc. ACL"}

@string{naacl = "Proc. NAACL"}

@inproceedings{tian2025espnet,
  title={{ESPnet}-{SpeechLM}: An open speech language model toolkit},
  author={Tian, Jinchuan and Shi, Jiatong and Chen, William and Arora, Siddhant and Masuyama, Yoshiki and Maekaku, Takashi and Wu, Yihan and Peng, Junyi and Bharadwaj, Shikhar and Zhao, Yiwen and others},
  booktitle=naacl,
  year={2025}
}

@article{zeghidour2021soundstream,
    author = {Zeghidour, Neil and Luebs, Alejandro and Omran, Ahmed and Skoglund, Jan and Tagliasacchi, Marco},
    title = {{SoundStream}: An End-to-End Neural Audio Codec},
    year = {2021},
    journal = {IEEE/ACM Transactions on Audio, Speech, and Language Processing},
    pages = {495--507},
}

@inproceedings{shi2024versa,
  title={{VERSA}: A Versatile Evaluation Toolkit for Speech, Audio, and Music},
  author={Shi, Jiatong and Shim, Hye-jin and Tian, Jinchuan and Arora, Siddhant and Wu, Haibin and Petermann, Darius and Yip, Jia Qi and Zhang, You and Tang, Yuxun and Zhang, Wangyou and others},
  booktitle=naacl,
  year={2025}
}

@article{wu2024towards,
  title={Towards audio language modeling-an overview},
  author={Wu, Haibin and Chen, Xuanjun and Lin, Yi-Cheng and Chang, Kai-wei and Chung, Ho-Lam and Liu, Alexander H. and Lee, {Hung-yi}},
  journal={arXiv preprint arXiv:2402.13236},
  year={2024}
}

@inproceedings{petermann2021harp,
  title={Harp-net: Hyper-autoencoded reconstruction propagation for scalable neural audio coding},
  author={Petermann, Darius and Beack, Seungkwon and Kim, Minje},
  booktitle=waspaa,
  pages={316--320},
  year={2021},
  organization={IEEE}
}

@article{Ji2024LanguageCodecRT,
  title={Language-{Codec}: Reducing the Gaps Between Discrete Codec Representation and Speech Language Models},
  author={Shengpeng Ji and Minghui Fang and Ziyue Jiang and Rongjie Huang and Jialung Zuo and Shulei Wang and Zhou Zhao},
  journal={arXiv preprint arXiv:2402.12208},
  year={2024},
}

@article{Ahn2024HILCodecHA,
  title={{HILCodec}: High-Fidelity and Lightweight Neural Audio Codec},
  author={Ahn, Sung Hwan and Woo, Beom Jun and Han, Mingrui and Moon, Chan Yeong and Kim, Nam Soo},
  journal={IEEE Journal of Selected Topics in Signal Processing},
  year={2024},
  volume={18},
  pages={1517-1530},
}

@article{Wu2024TS3CodecTS,
  title={{TS3-Codec}: Transformer-Based Simple Streaming Single Codec},
  author={Wu, Haibin and Kanda, Naoyuki and Eskimez, Sefik Emre and Li, Jinyu},
  journal={arXiv preprint arXiv:2411.18803},
  year={2024}
}

@inproceedings{Guo2024LSCodecLA,
  title={{LSCodec}: Low-Bitrate and Speaker-Decoupled Discrete Speech Codec},
  author={Yiwei Guo and Zhihan Li and Chenpeng Du and Hankun Wang and Xie Chen and Kai Yu},
  booktitle=interspeech,
  year={2025},
}

@article{Jiang2022DisentangledFL,
  title={Disentangled Feature Learning for Real-Time Neural Speech Coding},
  author={Xue Jiang and Xiulian Peng and Yuan Zhang and Yan Lu},
  journal=icassp, 
  year={2022},
  pages={1-5},
  organization={IEEE},
}

@inproceedings{kumar2023high,
    title={High-Fidelity Audio Compression with Improved {RVQGAN}},
    author={Rithesh Kumar and Prem Seetharaman and Alejandro Luebs and Ishaan Kumar and Kundan Kumar},
    booktitle=neurips,
    year={2023},
}

@inproceedings{zhang2023speechtokenizer,
    title={{SpeechTokenizer}: Unified speech tokenizer for speech large language models},
    author={Zhang, Xin and Zhang, Dong and Li, Shimin and Zhou, Yaqian and Qiu, Xipeng},
    booktitle=iclr,
    year={2024}
}

@inproceedings{wu2023audiodec,
    author={Wu, Yi-Chiao and Gebru, Israel D. and Marković, Dejan and Richard, Alexander},
    booktitle=icassp,
    title={Audiodec: An Open-Source Streaming High-Fidelity Neural Audio Codec}, 
    year={2023},
    pages={1--5},
    organization={IEEE},
}

@article{yang2023hifi,
    title={{HiFi-Codec}: Group-residual Vector quantization for High Fidelity Audio Codec},
    author={Dongchao Yang and Songxiang Liu and Rongjie Huang and Jinchuan Tian and Chao Weng and Yuexian Zou},
    journal={arXiv preprint arXiv:2305.02765},
    year={2023}
}

@article{du2023funcodec,
    title={{FunCodec}: A Fundamental, Reproducible and Integrable Open-source Toolkit for Neural Speech Codec},
    author={Zhihao Du and Shiliang Zhang and Kai Hu and Siqi Zheng},
    year={2023},
    journal={arXiv preprint arXiv:2309.07405},
}

@article{defossez2024moshi,
  title={Moshi: a speech-text foundation model for real-time dialogue},
  author={D{\'e}fossez, Alexandre and Mazar{\'e}, Laurent and Orsini, Manu and Royer, Am{\'e}lie and P{\'e}rez, Patrick and J{\'e}gou, Herv{\'e} and Grave, Edouard and Zeghidour, Neil},
  journal={arXiv preprint arXiv:2410.00037},
  year={2024}
}

@inproceedings{wu2024codec,
  title={{Codec-SUPERB}: An In-Depth Analysis of Sound Codec Models},
  author={Wu, Haibin and Chung, Ho-Lam and Lin, Yi-Cheng and Wu, Yuan-Kuei and Chen, Xuanjun and Pai, Yu-Chi and Wang, Hsiu-Hsuan and Chang, Kai-Wei and Liu, Alexander H and Lee, Hung-yi},
  booktitle = acl,
  year={2024}
}

@inproceedings{wichern2019wham,
    title={{WHAM!}: Extending Speech Separation to Noisy Environments}, 
    author={Gordon Wichern and others},
    year={2019},
    booktitle=interspeech, 
}

@inproceedings{1360861705599880960,
    author="Takaaki, Saeki and others",
    title="{{UTMOS}: {UTokyo}-{SaruLab} System for {VoiceMOS} Challenge 2022}",
    booktitle=interspeech,
    year="2022",
    pages={4521--4525},
}

@article{
defossez2022high,
title={High Fidelity Neural Audio Compression},
author={Alexandre D{\'e}fossez and Jade Copet and Gabriel Synnaeve and Yossi Adi},
journal={Transactions on Machine Learning Research},
issn={2835-8856},
year={2023},
url={https://openreview.net/forum?id=ivCd8z8zR2},
note={Featured Certification, Reproducibility Certification}
}

@article{dubey2024icassp,
  title={{ICASSP} 2023 deep noise suppression challenge},
  author={Dubey, Harishchandra and Aazami, Ashkan and Gopal, Vishak and Naderi, Babak and Braun, Sebastian and Cutler, Ross and Ju, Alex and Zohourian, Mehdi and Tang, Min and Golestaneh, Mehrsa and others},
  journal={IEEE Open Journal of Signal Processing},
  year={2024},
}

@inproceedings{gemmeke2017audio,
  title={{Audio Set}: An ontology and human-labeled dataset for audio events},
  author={Gemmeke, Jort F and Ellis, Daniel PW and Freedman, Dylan and Jansen, Aren and Lawrence, Wade and Moore, R Channing and Plakal, Manoj and Ritter, Marvin},
  booktitle=icassp, 
  pages={776--780},
  year={2017},
  organization={IEEE},
}

@inproceedings{reddy2022dnsmos,
    title={{DNSMOS P.835}: A non-intrusive perceptual objective speech quality metric to evaluate noise suppressors},
    author={Reddy, Chandan KA and Gopal, Vishak and Cutler, Ross},
    booktitle=icassp,
    year={2022}, 
    organization={IEEE},
}

@article{xin2024bigcodec,
  title={{BigCodec}: Pushing the limits of low-bitrate neural speech codec},
  author={Xin, Detai and Tan, Xu and Takamichi, Shinnosuke and Saruwatari, Hiroshi},
  journal={arXiv preprint arXiv:2409.05377},
  year={2024}
}

@article{siuzdak2024snac,
  title={{SNAC}: Multi-scale neural audio codec},
  author={Siuzdak, Hubert and Gr{\"o}tschla, Florian and Lanzend{\"o}rfer, Luca A},
  journal={arXiv preprint arXiv:2410.14411},
  year={2024}
}

@article{langman2024spectral,
  title={Spectral codecs: Spectrogram-based audio codecs for high quality speech synthesis},
  author={Langman, Ryan and Juki{\'c}, Ante and Dhawan, Kunal and Koluguri, Nithin Rao and Ginsburg, Boris},
  journal={arXiv preprint arXiv:2406.05298},
  year={2024}
}

@inproceedings{gu2024esc,
  title={{ESC}: Efficient speech coding with cross-scale residual vector quantized transformers},
  author={Gu, Yuzhe and Diao, Enmao},
  booktitle=emnlp,
  year={2024}
}

@inproceedings{messica2024nast,
  title={{NAST}: Noise aware speech tokenization for speech language models},
  author={Messica, Shoval and Adi, Yossi},
  booktitle=interspeech,
  year={2024}
}

@inproceedings{casanova2024lfsc,
  title={Low frame-rate speech codec: a codec designed for fast high-quality speech {LLM} training and inference},
  author={Casanova, Edresson and Langman, Ryan and Neekhara, Paarth and Hussain, Shehzeen and Li, Jason and Ghosh, Subhankar and Juki{\'c}, Ante and Lee, Sang-gil},
  booktitle=icassp,
  year={2025},
  organization={IEEE}
}

@inproceedings{
parker2024scaling,
title={Scaling Transformers for Low-Bitrate High-Quality Speech Coding},
author={Julian D Parker and Anton Smirnov and Jordi Pons and CJ Carr and Zack Zukowski and Zach Evans and Xubo Liu},
booktitle=icml,
year={2025},
url={https://openreview.net/forum?id=4YpMrGfldX}
}

@article{ai2024apcodec,
  title={{APCodec}: A neural audio codec with parallel amplitude and phase spectrum encoding and decoding},
  author={Ai, Yang and Jiang, Xiao-Hang and Lu, Ye-Xin and Du, Hui-Peng and Ling, Zhen-Hua},
  journal={IEEE/ACM Transactions on Audio, Speech, and Language Processing},
  year={2024},
  publisher={IEEE}
}

@inproceedings{shi2024espnet,
  title={{ESPnet-Codec}: Comprehensive training and evaluation of neural codecs for audio, music, and speech},
  author={Shi, Jiatong and Tian, Jinchuan and Wu, Yihan and Jung, Jee-weon and Yip, Jia Qi and Masuyama, Yoshiki and Chen, William and Wu, Yuning and Tang, Yuxun and Baali, Massa and others},
  booktitle=slt,
  pages={562--569},
  year={2024},
  organization={IEEE}
}

@article{zen2019libritts,
  title={Libri{TTS}: A Corpus Derived from LibriSpeech for Text-to-Speech},
  author={Zen, Heiga and Dang, Viet and Clark, Rob and Zhang, Yu and Weiss, Ron J and Jia, Ye and Chen, Zhifeng and Wu, Yonghui},
  journal=interspeech,
  year={2019},
  publisher={ISCA}
}

@INPROCEEDINGS{7178964,
  author={Panayotov, Vassil and Chen, Guoguo and Povey, Daniel and Khudanpur, Sanjeev},
  booktitle=icassp, 
  title={Librispeech: An {ASR} corpus based on public domain audio books}, 
  year={2015},
  volume={},
  number={},
  pages={5206-5210},
  organization={IEEE},
}

@inproceedings{yang2024generative,
  title={Generative de-quantization for neural speech codec via latent diffusion},
  author={Yang, Haici and Jang, Inseon and Kim, Minje},
  booktitle=icassp,
  pages={1251--1255},
  year={2024},
  organization={IEEE}
}

@article{cui2024recent,
  title={Recent advances in speech language models: A survey},
  author={Cui, Wenqian and Yu, Dianzhi and Jiao, Xiaoqi and Meng, Ziqiao and Zhang, Guangyan and Wang, Qichao and Guo, Yiwen and King, Irwin},
  journal={arXiv preprint arXiv:2410.03751},
  year={2024}
}

@inproceedings{radford2023robust,
  title={Robust speech recognition via large-scale weak supervision},
  author={Radford, Alec and Kim, Jong Wook and Xu, Tao and Brockman, Greg and McLeavey, Christine and Sutskever, Ilya},
  booktitle=icml,
  pages={28492--28518},
  year={2023},
  organization={PMLR}
}

@inproceedings{
yang2024uniaudio,
title={{UniAudio} 1.5: Large Language Model-Driven Audio Codec is A Few-Shot Audio Task Learner},
author={Dongchao Yang and Haohan Guo and Yuanyuan Wang and Rongjie Huang and Xiang Li and Xu Tan and Xixin Wu and Helen M. Meng},
booktitle=neurips,
year={2024},
url={https://openreview.net/forum?id=NGrINZyZKk}
}

@article{copet2024musicgen,
  title={Simple and controllable music generation},
  author={Copet, Jade and Kreuk, Felix and Gat, Itai and Remez, Tal and Kant, David and Synnaeve, Gabriel and Adi, Yossi and D{\'e}fossez, Alexandre},
  journal=neurips,
  volume={36},
  pages={47704--47720},
  year={2023}
}

@article{ji2024wavtokenizer,
  title={{WavTokenizer}: an Efficient Acoustic Discrete Codec Tokenizer for Audio Language Modeling},
  author={Ji, Shengpeng and Jiang, Ziyue and Cheng, Xize and Chen, Yifu and Fang, Minghui and Zuo, Jialong and Yang, Qian and Li, Ruiqi and Zhang, Ziang and Yang, Xiaoda and others},
  journal={CoRR},
  year={2024}
}

@article{guo2025recent,
  title={Recent Advances in Discrete Speech Tokens: A Review},
  author={Guo, Yiwei and Li, Zhihan and Wang, Hankun and Li, Bohan and Shao, Chongtian and Zhang, Hanglei and Du, Chenpeng and Chen, Xie and Liu, Shujie and Yu, Kai},
  journal={arXiv preprint arXiv:2502.06490},
  year={2025}
}

@article{du2024cosyvoiceAS,
  title={{CosyVoice}: A Scalable Multilingual Zero-shot Text-to-speech Synthesizer based on Supervised Semantic Tokens},
  author={Du, Zhihao and Chen, Qian and Zhang, Shiliang and Hu, Kai and Lu, Heng and Yang, Yexin and Hu, Hangrui and Zheng, Siqi and Gu, Yue and Ma, Ziyang and others},
  journal={arXiv preprint arXiv:2407.05407},
  year={2024}
}

@article{du2024cosyvoice,
  title={{CosyVoice} 2: Scalable streaming speech synthesis with large language models},
  author={Du, Zhihao and Wang, Yuxuan and Chen, Qian and Shi, Xian and Lv, Xiang and Zhao, Tianyu and Gao, Zhifu and Yang, Yexin and Gao, Changfeng and Wang, Hui and others},
  journal={arXiv preprint arXiv:2412.10117},
  year={2024}
}

@article{painter2000perceptual,
  title={Perceptual coding of digital audio},
  author={Painter, Ted and Spanias, Andreas},
  journal={Proceedings of the IEEE},
  volume={88},
  number={4},
  pages={451--515},
  year={2000},
  publisher={IEEE}
}

@inproceedings{ju2024naturalspeech,
  title={{NaturalSpeech} 3: Zero-Shot Speech Synthesis with Factorized Codec and Diffusion Models},
  author={Ju, Zeqian and Wang, Yuancheng and Shen, Kai and Tan, Xu and Xin, Detai and Yang, Dongchao and Liu, Eric and Leng, Yichong and Song, Kaitao and Tang, Siliang and others},
  booktitle=icml,
  pages={22605--22623},
  year={2024},
  organization={PMLR}
}

@article{painter2002perceptual,
  title={Perceptual coding of digital audio},
  author={Painter, Ted and Spanias, Andreas},
  journal={Proceedings of the IEEE},
  volume={88},
  number={4},
  pages={451--515},
  year={2002},
  publisher={IEEE}
}

@article{bessette2003adaptive,
  title={The adaptive multirate wideband speech codec ({AMR-WB})},
  author={Bessette, Bruno and Salami, Redwan and Lefebvre, Roch and Jelinek, Milan and Rotola-Pukkila, Jani and Vainio, Janne and Mikkola, Hannu and Jarvinen, Kari},
  journal={IEEE transactions on speech and audio processing},
  volume={10},
  number={8},
  pages={620--636},
  year={2003},
  publisher={IEEE}
}

@misc{opuscodec,
  author       = {Jean-Marc Valin and Koen Vos and Timothy B. Terriberry},
  title        = {Opus Codec: The standard for interactive speech and audio transmission over the {Internet}},
  howpublished = {https://opus-codec.org},
  year         = {2012},
  note         = {Accessed: 2025-05-27}
}

@article{barnes1996advances,
  title={Advances in residual vector quantization: A review},
  author={Barnes, Christopher F and Rizvi, Syed A and Nasrabadi, Nasser M},
  journal={IEEE transactions on image processing},
  volume={5},
  number={2},
  pages={226--262},
  year={1996},
  publisher={IEEE}
}

@inproceedings{tian2024effects,
  title={On the Effects of Heterogeneous Data Sources on Speech-to-Text Foundation Models},
  author={Tian, Jinchuan and Peng, Yifan and Chen, William and Choi, Kwanghee and Livescu, Karen and Watanabe, Shinji},
  booktitle={Proc. Interspeech 2024},
  pages={3959--3963},
  year={2024}
}

@inproceedings{ardila2020common,
  title={Common Voice: A Massively-Multilingual Speech Corpus},
  author={Ardila, Rosana and Branson, Megan and Davis, Kelly and Kohler, Michael and Meyer, Josh and Henretty, Michael and Morais, Reuben and Saunders, Lindsay and Tyers, Francis and Weber, Gregor},
  booktitle={Proceedings of the Twelfth Language Resources and Evaluation Conference},
  pages={4218--4222},
  year={2020}
}

@inproceedings{zhang2024urgent,
  title={{URGENT} Challenge: Universality, Robustness, and Generalizability For Speech Enhancement},
  author={Zhang, Wangyou and Scheibler, Robin and Saijo, Kohei and Cornell, Samuele and Li, Chenda and Ni, Zhaoheng and Pirklbauer, Jan and Sach, Marvin and Watanabe, Shinji and Fingscheidt, Tim and others},
  booktitle={Proc. Interspeech 2024},
  pages={4868--4872},
  year={2024}
}

@inproceedings{fonseca2017freesound,
  title={Freesound Datasets: A Platform for the Creation of Open Audio Datasets.},
  author={Fonseca, Eduardo and Pons, Jordi and Favory, Xavier and Font, Frederic and Bogdanov, Dmitry and Ferraro, Andres and Oramas, Sergio and Porter, Alastair and Serra, Xavier},
  booktitle={Proc. ISMIR},
  pages={486--493},
  year={2017}
}

@inproceedings{paul1992design,
  title={The design for the wall street journal-based {CSR} corpus},
  author={Paul, Douglas B and Baker, Janet M},
  booktitle={Proceedings of the workshop on Speech and Natural Language},
  pages={357--362},
  year={1992}
}

@article{wang2023tf,
  title={{TF-GridNet}: Integrating full-and sub-band modeling for speech separation},
  author={Wang, Zhong-Qiu and Cornell, Samuele and Choi, Shukjae and Lee, Younglo and Kim, Byeong-Yeol and Watanabe, Shinji},
  journal={IEEE/ACM Transactions on Audio, Speech, and Language Processing},
  volume={31},
  pages={3221--3236},
  year={2023},
  publisher={IEEE}
}

@inproceedings{jung2024espnet,
  title={{ESPnet-SPK}: full pipeline speaker embedding toolkit with reproducible recipes, self-supervised front-ends, and off-the-shelf models},
  author={Jung, Jee-weon and Zhang, Wangyou and Shi, Jiatong and Aldeneh, Zakaria and Higuchi, Takuya and Gichamba, Alex and Theobald, Barry-John and Hussen Abdelaziz, Ahmed and Watanabe, Shinji},
  booktitle={Proc. Interspeech 2024},
  pages={4278--4282},
  year={2024}
}

@article{hines2015visqol,
  title={{ViSQOL}: an objective speech quality model},
  author={Hines, Andrew and Skoglund, Jan and Kokaram, Anil C and Harte, Naomi},
  journal={EURASIP Journal on Audio, Speech, and Music Processing},
  volume={2015},
  pages={1--18},
  year={2015},
  publisher={Springer}
}

@inproceedings{johnston1988estimation,
  title={Estimation of perceptual entropy using noise masking criteria},
  author={Johnston, James D},
  booktitle={Proc. ICASSP},
  pages={2524--2527},
  year={1988},
  organization={IEEE}
}

@inproceedings{shi2021sequence,
  title={Sequence-to-sequence singing voice synthesis with perceptual entropy loss},
  author={Shi, Jiatong and Guo, Shuai and Huo, Nan and Zhang, Yuekai and Jin, Qin},
  booktitle={Proc. ICASSP},
  pages={76--80},
  year={2021},
  organization={IEEE}
}

@article{van2017neural,
  title={Neural discrete representation learning},
  author={Van Den Oord, Aaron and Vinyals, Oriol and others},
  journal={Advances in neural information processing systems},
  volume={30},
  year={2017}
}

@article{zhang2024scale,
  title={Scale This, Not That: Investigating Key Dataset Attributes for Efficient Speech Enhancement Scaling},
  author={Zhang, Leying and Zhang, Wangyou and Li, Chenda and Qian, Yanmin},
  journal={arXiv preprint arXiv:2412.14890},
  year={2024}
}

\end{document}